# Performance of dopamine modified $0.5(Ba_{0.7}Ca_{0.3})TiO_3$-$0.5Ba(Zr_{0.2}Ti_{0.8})O_3$ filler in PVDF nanocomposite as flexible energy storage and harvester

**Chhavi Mitharwal[1], Geetanjali[1], Shilpa Malhotra[1], Manish Kumar Srivastava[1], Surya Mohan Gupta[2,3], Chandra Mohan Singh Negi[1], Epsita Kar[1], Ajit R Kulkarni[4], Supratim Mitra[*1]**

[1] School of Physical Sciences, Banasthali Vidyapith, Banasthali 304022, India

[2] Homi Bhabha National Institute, Anushaktinagar, Mumbai 400094, India

[3] Laser Materials Section, Raja Ramanna Centre for Advanced Technology, Indore 452013, India

[4] Department of Metallurgical Engineering and Materials Science, Indian Institute of Technology Bombay, Mumbai 400076, India

[*] Corresponding author's email: supratimmitra@banasthali.in (Supratim Mitra)

## ABSTRACT

We demonstrate the potential of dopamine modified $0.5(Ba_{0.7}Ca_{0.3})TiO_3$-$0.5Ba(Zr_{0.2}Ti_{0.8})O_3$ filler incorporated poly-vinylidene fluoride (PVDF) composite prepared by solution cast method as both flexible energy storage and harvesting devices. The introduction of dopamine in filler surface functionalization acts as bridging elements between filler and polymer matrix and results in a better filler dispersion and an improved dielectric loss tangent (<0.02) along with dielectric permittivity ranges from 9 to 34 which is favorable for both energy harvesting and storage. Additionally, a significantly low DC conductivity ($< 10^{-9}$ ohm$^{-1}$cm$^{-1}$) for all composites was achieved leading to an improved breakdown strength and charge accumulation capability. Maximum breakdown strength of 134 KV/mm and corresponding energy storage density 0.72 J/cm$^3$ were obtained from the filler content 10 weight%. The improved energy harvesting performance was characterized by obtaining a maximum piezoelectric charge constant ($d_{33}$) = 78 pC/N, and output voltage ($V_{out}$) = 0.84 V along with maximum power density of 3.46 µW/cm$^3$ for the filler content of 10 wt%. Thus, the results show $0.5(Ba_{0.7}Ca_{0.3})TiO_3$-$0.5Ba(Zr_{0.2}Ti_{0.8})O_3$/PVDF composite has the potential for energy storage and harvesting applications simultaneously that can significantly suppress the excess energy loss arises while utilizing different material.

## 1. Introduction

In recent years, due to the advancement of smart, compact, and lightweight electronic devices operate under low power, sustainable energy sources are extremely important for the extended activity of such micro-devices. For the last decades, energy harvesters based on piezoelectricity from mechanical vibration are explored extensively for its functionality in energy technologies [1-6]. The utilization of piezoelectric materials for energy harvesting even from the irregular ambient vibrations that exist in the surrounding systems/structures is proved to be one among the foremost approachable mode for energy harvesting. Likewise, as the energy generated from these systems is unsteady and cannot be employed directly as a power source, a storage element either in the course of a capacitor or battery is normally needed. So, besides energy harvesting, energy storage devices have also become equally significant in energy technologies. Recently, electrostatic capacitors have gained much consideration by the researchers due to its high power density, large operating voltage, small leakage current, and fast charging/discharging time [7-12]. Thus, it can be realized that the improvement in the energy density of dielectric capacitors can lead to a miniaturized electronic storage system which might be an alternate for supercapacitor and even batteries. Furthermore, this can also promote integration of the energy harvester with a storage device to reduce undesired loss of energy because of the power management circuit in between and would be beneficial if the materials used in both are the same.

Amongst the many piezoelectric materials that have been investigated recently [13-16], ceramics based on barium titanate ($BaTiO_3$) [17-22], lead zirconate titanate (PZT) [23, 24] and polymers based on poly (vinylidene fluoride) (PVDF) [25, 26] are considered as potential candidates for the energy harvesting as a result of their excellent piezoelectric properties (i.e., high piezoelectric charge constant, $d_{33}$) and flexibility respectively. Besides, ferroelectric polymers have higher breakdown strength and low dielectric loss, whereas piezoelectric ceramics have high dielectric permittivity, which is considered as key factors for achieving excellent energy storage density given as,

$$U_{st} = \frac{1}{2}\varepsilon_0\varepsilon_r E^2 \qquad (1)$$

where $\varepsilon_0$ is the permittivity of free space and $\varepsilon_r$ is the relative permittivity (or dielectric constant) of the dielectric material and $E$ is limited by the breakdown field $E_{BD}$ [27]. Although $BaTiO_3$ and PZT possess high $d_{33}$ = 100, and 200 pC/N respectively, they are brittle while PVDF being highly flexible could not be used largely in energy harvester alone due to lack in piezoelectric properties ($d_{33}$ = 33 pC/N). Therefore, the approach has always been the preparation of a composite via the synergistic impact of both the component to obtain optimized electrical and mechanical properties for flexible energy storage and harvester [20, 28-32]. Ceramic fillers in nanostructured form have been the primary choice to obtain a polymer nanocomposite as it can induce the electro-active β-phase of PVDF and also retain the same β-phase during solidification [25, 33]. For this, $BaTiO_3$ and PZT nanoparticles (NPs) are extensively used, however, $BaTiO_3$ NPs are preferred over PZT NPs due to its non-toxic and biocompatible nature [33, 34]. Further, the homogeneous dispersion of these nanostructured fillers in the polymer matrix becomes difficult due to different surface characteristics. Therefore, to get a better compatibility between fillers and matrix, surface of the filler particles are functionalized which turns into a key factor to achieve a high-performance nanocomposite. Dopamine- a bio-inspired molecule has been used broadly for surface functionalization of several inorganic compounds including $BaTiO_3$ [35-37]. A hydrogen bond between fillers and dopamine and a reaction between $NH_2$ of dopamine with C-F group of PVDF polymer provides the linkage between fillers and matrix through dopamine.

In the present study, we have selected $0.5(Ba_{0.7}Ca_{0.3})TiO_3$-$0.5Ba(Zr_{0.2}Ti_{0.8})O_3$ [BCT-BZT] fillers known to have high $d_{33}$ = 620 pC/N, high $\varepsilon_r$ = 3200 compared to $BaTiO_3$ ($d_{33}$ = 100 pC/N and high $\varepsilon_r$ = 1200) [38] and along with PVDF as polymer matrix for the preparation of the composite. Further, to obtain improved compatibility among filler and polymer network as well as uniform distribution of fillers, BCT-BZT nanoparticles were functionalized using dopamine. Thus, the influence of dopamine modified fillers on dielectric properties, energy harvesting, and storage performances are systematically studied.

## 2. Experimental

For the preparation of BCT-BZT/PVDF composite, the filler nano particles $0.5(Ba_{0.7}Ca_{0.3})TiO_3$-$0.5Ba(Zr_{0.2}Ti_{0.8})O_3$ [BCT-BZT] were prepared using sol-gel method (details are given in Supplementary) and further functionalized by dopamine (referred to as WDA). Another batch of BCT-BZT/PVDF composite were also prepared without modifying the fillers (referred as WODA). A different weight fractions (10-60 wt%) of fillers were used in solution cast method to form free-standing films. Firstly, poly (vinylidene fluoride) (PVDF) (Sigma Aldrich, USA) was dissolved in Dimethyl sulfoxide (DMSO) (Sigma Aldrich, USA) at 80 °C and then filler particles were added slowly and stirred continuously for 12h followed by sonication for 90 min. Then, the homogeneous slurry was cast on a clean petri dish and kept at 60 °C for solvent evaporation. The composite films were peeled off at room temperature and thickness ~ 50 µm were obtained. The phase formation of the composite films was also confirmed using an X-ray diffractometer (Bruker D8 Advance, Germany) and morphology was studied using a scanning electron microscope (SEM) (Tescan MIRA 3 MUG FEG, Czech Republic). Further, for electrical measurements, composite films were sputter-coated with Al on both sides. Dielectric measurements were carried out using a precision impedance analyzer (Wayne Kerr, 6510B, UK), the breakdown strength (BDS) was measured using a DC power supply (Ionics, India) test setup. To test energy harvesting performance based on human motion, the composites were cut into a rectangular shape having an active area of 4.5×1.5 cm² and mechanically excited under the periodic beating with the fist. The open-circuit voltage ($V_{oc}$) and short-circuit current density ($J_{sc}$) were measured by a digital oscilloscope (SDS 1052DL, Siglent Technologies, Netherlands). To calculate maximum power density, a load resistance ($R_L$) varied from 1 kΩ to $10^6$ kΩ was attached in parallel to output.

## 3. Result and discussion

### 3.1. Structure Analysis

The XRD patterns of BCT-BZT/PVDF composites for WODA and WDA are presented in Fig. 1(a) and 1(b) respectively. The pseudo-cubic peaks associated to BCT-BZT powders and α-PVDF peak at 2θ = 18.3º for (020), a broad β-PVDF peak at 2θ = 20.26º for (100)/(200) in WODA are identified for the filler content up to 40 wt%. In addition, the β-PVDF peak at 2θ = 18.5º for (020) was also observed in WDA for the filler content up to 40 wt% along with the same peaks observed in WODA [25]. This suggests that the dopamine modified fillers induce the β-phase of PVDF without affecting the crystal structure of BCT-BZT filler particles. The decrease in intensities of PVDF peaks with an increase in filler content above 40 wt% in both WODA and WDA implies the dominance of the BCT-BZT filler phase in the composites.

### 3.2. Microstructure analysis

The SEM micrographs of the top surface of as-cast WDA and WODA composite films are depicted in Figs. 2(a-c) and (d-f), respectively, for filler content of 10 wt% (as representative composition). Overall, a dense microstructure with uniform grains free from any spherulite, a typical characteristic of α-phase of PVDF, has been observed as shown in Fig. 2(a) and (d) [39-41]. A similar grain structure was also observed for all compositions, however, with varying grain sizes. The average grain size was measured using a statistical average method of the grain size measurements which was carried out over 50-100 grains using a software Image J. The variation of average grain size with filler content is shown in Fig. 2(g) and found to decrease with the increase in filler content for both WDA and WODA composite, while the average grain size of WDA is found to have a lower value than that of WODA. Meanwhile, the magnified images are presented in Fig. 2(c) and (f) for the WDA and WODA respectively, showed that WDA fillers are comparatively well dispersed in the polymer matrix, while larger size agglomerates of fillers are observed in WODA. Therefore, for further understanding the influence of dopamine modification of nanofillers on microstructure development and dispersion, a schematic is presented in Fig. 2(h). The microstructure is developed in the composite by nucleation and subsequent growth of the nuclei by the coalescence process very similar to the growth of metal film on non-wetting surfaces [42]. Here, each nanofillers act as nucleation centres and therefore, accelerate the nucleation process and finally grow like a grain. Therefore, as the filler content is increased, the density of the nucleation centre in the composite increases, which results in a greater number of grains within a given volume. This led to smaller grain size by hindering the grain growth as depicted in the schematic. In the meanwhile, after the post-nucleation stage, several nuclei coalescences by minimizing the surface free energy [42-44]. This results in the presence of many fillers within a single grain of the composite which is seen in the magnified images of Fig. 2(c) and (f). From these micrographs, an improved dispersion in WDA can be seen which arises due to the better compatibility among the filler and polymer matrix [35-37], that might be established by the formation of a hydrogen bond between filler and dopamine and on the other hand reaction between $NH_2$ of dopamine with C-F group of PVDF polymer as presented in Fig. 3 [35]. This chemical bridging between filler and matrix through dopamine results in better dispersion. Further, the composite with dense microstructure and smaller grain size is known to have a large breakdown strength which also results in high energy storage density [45, 46].

### 3.3. Dielectric analysis

The frequency-dependent (20 Hz – $10^7$ Hz) dielectric permittivity, loss tangent, and AC conductivity measured at room temperature for different filler content of BCT-BZT/PVDF composites for both WODA and WDA modified fillers are presented in Fig. 4(a-f) (WDA: open symbol and WODA: solid symbol). As seen in Fig. 4(a) and (b), the dielectric permittivity for both WODA and WDA is almost frequency independent up to $10^6$ Hz and then decreases gradually up to $10^7$ Hz for all composites. At higher frequencies, the frequency of switching of electrical dipoles could not follow the applied frequency and thereby dielectric permittivity decreases. Meanwhile, the dielectric loss tangent continues to stay constant (< 0.08) for all other composites in the frequency range from $10^2$ to $10^5$ Hz as seen in Fig. 4(c) and (d) and in their respective insets. However, slightly higher values are observed in the lower frequencies ($<10^2$) that can be associated with Maxwell-Wagner interfacial polarization. Also, at frequencies above $10^5$ Hz, the loss tangent increases sharply to form the loss peak, which could be attributed to α-relaxation and associated glass transition of the PVDF matrix [47-50].

Further, to look into the effect of dopamine modified BCT-BZT filler on dielectric properties and conductivity of BCT-BZT/PVDF composite for both WODA and WDA modified fillers, the dielectric permittivity, loss tangent measured at 1 kHz are depicted in Fig. 4(g) and (i) respectively. It is very clear from Fig. 4(g) that the dielectric permittivity in both WODA and WDA has increased with filler content and the values are higher in WDA composites. The dielectric permittivity value has increased from 6.5 for neat PVDF to 34.5 for 60 wt% in WDA composite. In addition, a significantly lower value (< 0.02) of the dielectric loss tangent for all WDA composites compared to neat PVDF (0.032) is seen in Fig. 4(h). Comparatively, a higher loss tangent for all compositions of WODA is observed than WDA composite. With the introduction of filler content having very high dielectric permittivity, the average electric field increased, leading to enhancement in the dielectric permittivity. Additionally, BCT-BZT fillers show low dielectric loss value while compared to pure PVDF, therefore the addition of filler in polymer matrix reduces the loss tangent value. Thus, the surface modification of BCT-BZT filler with dopamine plays a crucial role in the conjugation among filler and polymer matrix which results in improved permittivity and low loss tangent [35].

The AC conductivity of all compositions of WODA and WDA as seen in Fig. 4(e) and (f) respectively, show strong frequency dependence, which is governed by the equation,

$$\sigma_{AC}(\omega) = \sigma_{DC}(0) + A\omega^n \qquad (2)$$

where $\sigma_{DC}$ is DC conductivity, $\omega$ $(=2\pi f)$ is the angular frequency, $A$ is a constant, $n$ is an exponent lies in the range $0 < n < 1$ [51]. The DC conductivity was calculated from the linear fitting of equation (2) shown in Fig. 4(i) (inset shows the linear fitting of PVDF as an example) for WODA and WDA composites. The values of DC conductivity are found to increase with filler content for both WODA and WDA composites, however, comparatively a much lower value was observed for WDA for the same filler content. This could be explained by the fact that dopamine restricts the charge transportation between two adjacent filler particles that are created between the interface of filler and polymer matrix [52]. However, if the filler concentration is very high, i.e., interparticle distance becomes significantly small and there could be the possibility of formation of a conducting path [53]. A lower value of conductivity is always expected to have high breakdown strength, and that becomes a decisive factor in energy storage performance.

### 3.4. Energy storage performance

The improvement in the energy storage density is realized by further improving the breakdown strength (BDS) (Eq. 1), which is calculated from Weibull statistics, and given by the equation [54],

$$P(E) = 1- exp(-(E/E_b)^\beta) \qquad (3)$$

Where $P(E)$ is the cumulative probability of failure of experimental breakdown field $E$, $E_b$ is Weibull characteristics BDS where cumulative failure probability is 63.2%, i.e., $P(E_b) = 63.2\%$ and $\beta$ is the shape parameter representing the reliability of data. For each experimental $E$, $P(E)$ can be estimated from,

$$P = (i - 0.44)/(n + 0.25) \qquad (4)$$

Where $i$ is the $i^{th}$ rank of $E$ in the ascending order of BDS data and $n$ is the total no of data points. After simplifying equation (3) into a linear equation, $\ln(\ln(1-P)^{-1})$ vs $\ln(E)$ is plotted as seen in Fig. 5(a) and (b) for WODA and WDA, respectively. From the linear fitting, $E_b$ is calculated as shown in Fig. 5(c) where, the error bar represents the reliability of data from the shape parameter, β mentioned in Equation 3. A maximum value of $E_b = 134$ kV/mm was obtained for filler content 10 wt% of WDA composite which then decreased with increasing filler content. A similar trend of variation of $E_b$ with filler content was also observed in WODA however, the values of $E_b$ that obtained for all WDA composites are found larger than WODA composites. This is because the dopamine modified composites (WDA) have lower electrical conductivity and in both cases, as the filler content is increased, conductivity is found to increase as also seen in Fig. 4(i). The variation of energy density with filler content is calculated from Eqn. (1) and are depicted in Fig. 5(d). The maximum energy storage density, $U_{st} = 0.75$ J/cm$^3$ is obtained for 10 wt% WDA composite. Thus, it is seen that dopamine modified composites showed better energy storage performances.

**3.5. Energy harvesting performances**

Thus, to obtain the maximum output from energy harvester, 10 wt% WDA composite with highest energy density (0.75 J/cm$^3$) value was chosen for human-motion based energy harvesting performance as depicted in Fig. 6. The peak to peak open-circuit voltage ($V_{oc}$) which typically measured across a very high resistance (10 MΩ), and short-circuit current density ($J_{sc}$) measured across a very low resistance (1 kΩ) was obtained from the device of an active area of 4.5 × 1.6 cm$^2$ under the periodic beating with a fist as shown in Fig. 6 (a, b). A maximum peak-to-peak $V_{oc} \sim 30$ V and $J_{sc} \sim 120$ µA/cm$^2$ are observed during the pressing and releasing condition as presented in the inset of Fig. 6(a) and (b) respectively. This is a common approach of representing the output of energy harvesting devices, however, sometimes leads to an overestimation of output power density ($V_{oc} \times J_{sc}$) [55]. Therefore, to have a practical implementation of the device, the rectified output voltage ($V_{out}$) obtained using a rectifier circuit (Fig. 6(c)) and the corresponding power density of the device across an externally connected variable load resistance ($R_L$) was measured (Fig. 6(c)). The instantaneous power calculated from the equation, $P = V_{out}^2 / R_L$, and power density by using P/(area×thickness) was recorded for a range of $R_L$ varied from 1 kΩ to $10^6$ kΩ. A maximum power density of 3.46 µW/cm$^3$ was obtained at $R_L$ = $10^4$ kΩ and maximum $V_{out}$ = 0.84 V from the flexible energy harvester as depicted in the inset of Fig. 6(c). This shows the importance of calculating power across load resistance rather than using $V_{oc} \times J_{sc}$ as the later results in a large overestimation of the actual power generated. The charging and discharging capability of the output voltage is also evaluated using a capacitor of 10 µF as depicted in Fig. 6(d). The charging voltage is found to reach a maximum of 0.73 V in 18.04 sec, while it took almost 30 sec to discharge completely. The higher discharging time as compared to charging time shows the potential of BCT-BZT/PVDF composites as powering discharged batteries.

**Conclusions**

In summary, we have prepared 0.5(Ba$_{0.7}$Ca$_{0.3}$)TiO$_3$-0.5Ba(Zr$_{0.2}$Ti$_{0.8}$)O$_3$/PVDF composite for a wide range of filler content (10-60 wt%) via a simple solution casting method. The influence of surface modification by using dopamine have been systematically investigated for dielectric properties, energy storage, and harvesting characteristics. The results indicate that improved loss tangent (<0.02) along with dielectric permittivity ranges from 9 to 34 in dopamine modified composites is beneficial for energy storage performances. The DC conductivity in dopamine modified composites is reduced (< $10^{-9}$ ohm$^{-1}$cm$^{-1}$) by restricting the charge transport between adjacent filler particles and results in higher breakdown strength. A maximum breakdown strength of 134 KV/mm and corresponding energy storage density 0.72 J/cm$^3$ were obtained from the filler content 10 wt%. A reduction in DC conductivity also results in more accumulation of surface charges which showed a higher output voltage. The energy harvesting performance was characterized by obtaining output voltage ($V_{out}$) = 0.84 V along with maximum power density of 3.46 µW/cm$^3$ for the filler content of 10 wt%. Thus, the results confirm that 10 wt% dopamine modified 0.5(Ba$_{0.7}$Ca$_{0.3}$)TiO$_3$-0.5Ba(Zr$_{0.2}$Ti$_{0.8}$)O$_3$/PVDF composite showed both energy harvesting and storage capability and therefore might be a promising candidate for simultaneous energy storage and harvester by integrating into a single device.

**Acknowledgment:** The authors' SM, CM thank the support of Science and Engineering Board (SERB), Department of Science and Technology (DST), India for financial support under Early Career Research (ECR) grant (File No. 2016/ECR/000794/ES). Author SM would like to Thank Dr. Anjana Jain for scientific discussion on the problem and helping in improving the manuscript.

**Conflict of Interest Statement:** The authors declare that there is no conflict of interest.

**References**

[1]     K. Uchino, "Piezoelectric Energy Harvesting Systems—Essentials to Successful Developments," *Energy Technology,* vol. 6, pp. 829-848, 2018.
[2]     X. Wang, "Piezoelectric nanogenerators-Harvesting ambient mechanical energy at the nanometer scale," *Nano Energy,* vol. 1, pp. 13-24, 2012.
[3]     C. R. Bowen, H. A. Kim, P. M. Weaver, and S. Dunn, "Piezoelectric and ferroelectric materials and structures for energy harvesting applications," *Energy and Environmental Science,* vol. 7, pp. 25-44, 2014.
[4]     H. Li, C. Tian, and Z. D. Deng, "Energy harvesting from low frequency applications using piezoelectric materials," *Applied Physics Reviews,* vol. 1, 2014.
[5]     H. A. Sodano, D. J. Inman, and G. Park, "A review of power harvesting from vibration using piezoelectric materials," *Shock and Vibration Digest,* vol. 36, pp. 197-205, 2004.
[6]     C. Wan and C. R. Bowen, "Multiscale-structuring of polyvinylidene fluoride for energy harvesting: the impact of molecular-, micro- and macro-structure," *Journal of Materials Chemistry A,* vol. 5, pp. 3091-3128, 2017.
[7]     A. Bezryadin, A. Belkin, E. Ilin, M. Pak, E. V. Colla, and A. Hubler, "Large energy storage efficiency of the dielectric layer of graphene nanocapacitors," *Nanotechnology,* vol. 28, p. 495401, Dec 8 2017.
[8]     X. Hao, "A review on the dielectric materials for high energy-storage application," *Journal of Advanced Dielectrics,* vol. 03, p. 1330001, 2013/01/01 2013.
[9]     B. Chu, X. Zhou, K. Ren, B. Neese, M. Lin, Q. Wang*, et al.*, "A Dielectric Polymer with High Electric Energy Density and Fast Discharge Speed," *Science,* vol. 313, p. 334, 2006.
[10]    Z.-M. Dang, J.-K. Yuan, S.-H. Yao, and R.-J. Liao, "Flexible Nanodielectric Materials with High Permittivity for Power Energy Storage," *Advanced Materials,* vol. 25, pp. 6334-6365, 2013.

[11] Q. Li, G. Zhang, F. Liu, K. Han, M. R. Gadinski, C. Xiong, *et al.*, "Solution-processed ferroelectric terpolymer nanocomposites with high breakdown strength and energy density utilizing boron nitride nanosheets," *Energy & Environmental Science,* vol. 8, pp. 922-931, 2015.

[12] Prateek, V. K. Thakur, and R. K. Gupta, "Recent Progress on Ferroelectric Polymer-Based Nanocomposites for High Energy Density Capacitors: Synthesis, Dielectric Properties, and Future Aspects," *Chemical Reviews,* vol. 116, pp. 4260-4317, 2016/04/13 2016.

[13] J. Wu, D. Xiao, and J. Zhu, "Potassium-sodium niobate lead-free piezoelectric materials: past, present, and future of phase boundaries," *Chem Rev,* vol. 115, pp. 2559-95, Apr 8 2015.

[14] T. Zheng, J. Wu, D. Xiao, and J. Zhu, "Recent development in lead-free perovskite piezoelectric bulk materials," *Progress in Materials Science,* vol. 98, pp. 552-624, 2018.

[15] D. Xiao, "Progresses and Further Considerations on the Research of Perovskite Lead-Free Piezoelectric Ceramics," *Journal of Advanced Dielectrics,* vol. 01, pp. 33-40, 2012.

[16] S. Mitra, A. R. Kulkarni, and O. Prakash, "Diffuse phase transition and electrical properties of lead-free piezoelectric ($Li_xNa_{1-x}$)$NbO_3$ (0.04 ≤ x ≤ 0.20) ceramics near morphotropic phase boundary," *Journal of Applied Physics,* vol. 114, p. 064106, 2013.

[17] Z.-H. Lin, Y. Yang, J. M. Wu, Y. Liu, F. Zhang, and Z. L. Wang, "$BaTiO_3$ Nanotubes-Based Flexible and Transparent Nanogenerators," *The Journal of Physical Chemistry Letters,* vol. 3, pp. 3599-3604, 2012/12/06 2012.

[18] S. Mitra and A. R. Kulkarni, "Synthesis and Electrical Properties of New Lead-free (100−x)($Li_{0.12}Na_{0.88}$)$NbO_3$–x$BaTiO_3$ (0 ≤ x ≤ 40) Piezoelectric Ceramics," *Journal of the American Ceramic Society,* vol. 99, pp. 888-895, 2016/03/01 2016.

[19] K.-I. Park, M. Lee, Y. Liu, S. Moon, G.-T. Hwang, G. Zhu, *et al.*, "Flexible Nanocomposite Generator Made of $BaTiO_3$ Nanoparticles and Graphitic Carbons," *Advanced Materials,* vol. 24, pp. 2999-3004, 2012/06/12 2012.

[20] J. Nunes-Pereira, V. Sencadas, V. Correia, V. F. Cardoso, W. Han, J. G. Rocha, *et al.*, "Energy harvesting performance of $BaTiO_3$/poly(vinylidene fluoride–trifluoroethylene) spin coated nanocomposites," *Composites Part B: Engineering,* vol. 72, pp. 130-136, 2015.

[21] K.-i. Kakimoto, K. Fukata, and H. Ogawa, "Fabrication of fibrous $BaTiO_3$-reinforced PVDF composite sheet for transducer application," *Sensors and Actuators A: Physical,* vol. 200, pp. 21-25, 2013/10/01/ 2013.

[22] S. Mitra, T. Karthik, J. Kolte, R. Ade, N. Venkataramani, and A. R. Kulkarni, "Origin of enhanced piezoelectric properties and room temperature multiferroism in $MnO_2$ added 0.90($Li_{0.12}Na_{0.88}NbO_3$)-0.10$BaTiO_3$ ceramic," *Scripta Materialia,* vol. 149, pp. 134-138, 2018.

[23] B. Li, J. H. You, and Y.-J. Kim, "Low frequency acoustic energy harvesting using PZT piezoelectric plates in a straight tube resonator," *Smart Materials and Structures,* vol. 22, p. 055013, 2013/04/05 2013.

[24] S. Bai, Q. Xu, L. Gu, F. Ma, Y. Qin, and Z. L. Wang, "Single crystalline lead zirconate titanate (PZT) nano/micro-wire based self-powered UV sensor," *Nano Energy,* vol. 1, pp. 789-795, 2012/11/01/ 2012.

[25] P. Martins, A. C. Lopes, and S. Lanceros-Mendez, "Electroactive phases of poly(vinylidene fluoride): Determination, processing and applications," *Progress in Polymer Science,* vol. 39, pp. 683-706, 2014/04/01/ 2014.

[26] Z. Pi, J. Zhang, C. Wen, Z.-b. Zhang, and D. Wu, "Flexible piezoelectric nanogenerator made of poly(vinylidenefluoride-co-trifluoroethylene) (PVDF-TrFE) thin film," *Nano Energy,* vol. 7, pp. 33-41, 2014/07/01/ 2014.

[27] F. Liu, Q. Li, J. Cui, Z. Li, G. Yang, Y. Liu, *et al.*, "High-Energy-Density Dielectric Polymer Nanocomposites with Trilayered Architecture," *Advanced Functional Materials,* vol. 27, p. 1606292, 2017.

[28] T. Yamada, T. Ueda, and T. Kitayama, "Piezoelectricity of a high-content lead zirconate titanate/polymer composite," *Journal of Applied Physics,* vol. 53, pp. 4328-4332, 1982/06/01 1982.

[29] X. Jing, X. Shen, H. Song, and F. Song, "Magnetic and dielectric properties of barium ferrite fibers/poly(vinylidene fluoride) composite films," *Journal of Polymer Research,* vol. 18, pp. 2017-2021, 2011/11/01 2011.

[30] B. Dutta, E. Kar, N. Bose, and S. Mukherjee, "NiO@SiO2/PVDF: A Flexible Polymer Nanocomposite for a High Performance Human Body Motion-Based Energy Harvester and Tactile e-Skin Mechanosensor," *ACS Sustainable Chemistry & Engineering,* vol. 6, pp. 10505-10516, 2018/08/06 2018.

[31] E. Kar, N. Bose, B. Dutta, N. Mukherjee, and S. Mukherjee, "MWCNT@SiO2 Heterogeneous Nanofiller-Based Polymer Composites: A Single Key to the High-Performance Piezoelectric Nanogenerator and X-band Microwave Shield," *ACS Applied Nano Materials,* vol. 1, pp. 4005-4018, 2018/08/24 2018.

[32] E. Kar, N. Bose, B. Dutta, N. Mukherjee, and S. Mukherjee, "Ultraviolet- and Microwave-Protecting, Self-Cleaning e-Skin for Efficient Energy Harvesting and Tactile Mechanosensing," *ACS Applied Materials & Interfaces,* vol. 11, pp. 17501-17512, 2019/05/15 2019.

[33] S. Bodkhe, G. Turcot, F. P. Gosselin, and D. Therriault, "One-Step Solvent Evaporation-Assisted 3D Printing of Piezoelectric PVDF Nanocomposite Structures," *ACS Appl Mater Interfaces,* vol. 9, pp. 20833-20842, Jun 21 2017.

[34] H.-J. Ye, W.-Z. Shao, and L. Zhen, "Crystallization kinetics and phase transformation of poly(vinylidene fluoride) films incorporated with functionalized baTiO3 nanoparticles," *Journal of Applied Polymer Science,* vol. 129, pp. 2940-2949, 2013/09/05 2013.

[35] M.-F. Lin, V. K. Thakur, E. J. Tan, and P. S. Lee, "Surface functionalization of BaTiO3 nanoparticles and improved electrical properties of BaTiO3/polyvinylidene fluoride composite," *RSC Advances,* vol. 1, p. 576, 2011.

[36] Y. Xie, Y. Yu, Y. Feng, W. Jiang, and Z. Zhang, "Fabrication of Stretchable Nanocomposites with High Energy Density and Low Loss from Cross-Linked PVDF Filled with Poly(dopamine) Encapsulated BaTiO3," *ACS Appl Mater Interfaces,* vol. 9, pp. 2995-3005, Jan 25 2017.

[37] Y. Li, J. Yuan, J. Xue, F. Cai, F. Chen, and Q. Fu, "Towards suppressing loss tangent: Effect of polydopamine coating layers on dielectric properties of core–shell barium titanate filled polyvinylidene fluoride composites," *Composites Science and Technology,* vol. 118, pp. 198-206, 2015.

[38] W. Liu and X. Ren, "Large piezoelectric effect in Pb-free ceramics," *Phys Rev Lett,* vol. 103, p. 257602, Dec 18 2009.

[39] M. Tomizuka, A. Jain, J. Kumar S, D. R. Mahapatra, and H. H. Kumar, "Detailed studies on the formation of piezoelectric β-phase of PVDF at different hot-stretching conditions," vol. 7647, p. 76472C, 2010.

[40] A. Jain, J. S. Kumar, S. Srikanth, V. T. Rathod, and D. Roy Mahapatra, "Sensitivity of polyvinylidene fluoride films to mechanical vibration modes and impact after optimizing stretching conditions," *Polymer Engineering & Science,* vol. 53, pp. 707-715, 2013.

[41] A. Jain, S. J. Kumar, M. R. Kumar, A. S. Ganesh, and S. Srikanth, "PVDF-PZT Composite Films for Transducer Applications," *Mechanics of Advanced Materials and Structures,* vol. 21, pp. 181-186, 2013.

[42] M. Li, I. Katsouras, C. Piliego, G. Glasser, I. Lieberwirth, P. W. M. Blom*, et al.*, "Controlling the microstructure of poly(vinylidene-fluoride) (PVDF) thin films for microelectronics," *Journal of Materials Chemistry C,* vol. 1, p. 7695, 2013.

[43] S. F. Mendes, C. M. Costa, C. Caparros, V. Sencadas, and S. Lanceros-Méndez, "Effect of filler size and concentration on the structure and properties of poly(vinylidene fluoride)/BaTiO3 nanocomposites," *Journal of Materials Science,* vol. 47, pp. 1378-1388, 2012/02/01 2012.

[44] C.-W. Tang, B. Li, L. Sun, B. Lively, and W.-H. Zhong, "The effects of nanofillers, stretching and recrystallization on microstructure, phase transformation and dielectric properties in PVDF nanocomposites," *European Polymer Journal,* vol. 48, pp. 1062-1072, 2012.

[45] P. Kim, N. M. Doss, J. P. Tillotson, P. J. Hotchkiss, M.-J. Pan, S. R. Marder*, et al.*, "High Energy Density Nanocomposites Based on Surface-Modified BaTiO3 and a Ferroelectric Polymer," *ACS Nano,* vol. 3, pp. 2581-2592, 2009/09/22 2009.

[46] L. Yang, X. Kong, F. Li, H. Hao, Z. Cheng, H. Liu*, et al.*, "Perovskite lead-free dielectrics for energy storage applications," *Progress in Materials Science,* vol. 102, pp. 72-108, 2019/05/01/ 2019.

[47] S. Liu, S. Xue, W. Zhang, J. Zhai, and G. Chen, "Significantly enhanced dielectric property in PVDF nanocomposites flexible films through a small loading of surface-hydroxylated Ba0.6Sr0.4TiO3 nanotubes," *Journal of Materials Chemistry A,* vol. 2, pp. 18040-18046, 2014.

[48] B. Luo, X. Wang, Y. Wang, and L. Li, "Fabrication, characterization, properties and theoretical analysis of ceramic/PVDF composite flexible films with high dielectric constant and low dielectric loss," *Journal of Materials Chemistry A,* vol. 2, pp. 510-519, 2014.

[49] H. Luo, D. Zhang, C. Jiang, X. Yuan, C. Chen, and K. Zhou, "Improved Dielectric Properties and Energy Storage Density of Poly(vinylidene fluoride-co-hexafluoropropylene) Nanocomposite with Hydantoin Epoxy Resin Coated BaTiO3," *ACS Applied Materials & Interfaces,* vol. 7, pp. 8061-8069, 2015/04/22 2015.

[50] K. Yu, H. Wang, Y. Zhou, Y. Bai, and Y. Niu, "Enhanced dielectric properties of BaTiO3/poly(vinylidene fluoride) nanocomposites for energy storage applications," *Journal of Applied Physics,* vol. 113, p. 034105, 2013/01/21 2013.

[51] D. K. Pradhan, R. N. P. Choudhary, C. Rinaldi, and R. S. Katiyar, "Effect of Mn substitution on electrical and magnetic properties of Bi0.9La0.1FeO3," *Journal of Applied Physics,* vol. 106, p. 024102, 2009.

[52] Z. Wang, T. Wang, C. Wang, Y. Xiao, P. Jing, Y. Cui*, et al.*, "Poly(vinylidene fluoride) Flexible Nanocomposite Films with Dopamine-Coated Giant Dielectric Ceramic Nanopowders, Ba(Fe0.5Ta0.5)O3, for High Energy-Storage Density at Low Electric Field," *ACS Appl Mater Interfaces,* vol. 9, pp. 29130-29139, Aug 30 2017.

[53] G. Chen, X. Wang, J. Lin, W. Yang, H. Li, Y. Wen*, et al.*, "Nano-KTN@Ag/PVDF composite films with high permittivity and low dielectric loss by introduction of designed KTN/Ag core/shell nanoparticles," *Journal of Materials Chemistry C,* vol. 4, pp. 8070-8076, 2016.

[54] D. He, Y. Wang, X. Chen, and Y. Deng, "Core–shell structured BaTiO3@Al2O3 nanoparticles in polymer composites for dielectric loss suppression and breakdown strength enhancement," *Composites Part A: Applied Science and Manufacturing,* vol. 93, pp. 137-143, 2017/02/01/ 2017.

[55] J. Briscoe, N. Jalali, P. Woolliams, M. Stewart, P. M. Weaver, M. Cain*, et al.*, "Measurement techniques for piezoelectric nanogenerators," *Energy & Environmental Science,* vol. 6, p. 3035, 2013.

**Figure captions**

**Figure 1:** XRD patterns of BCT-BZT/PVDF composite films for (a) WODA and (b) WDA fillers.

**Figure 2:** SEM micrographs of BCT-BZT/PVDF composite films with filler content 10 wt% for (a-b) WDA and (c-d) WODA, respectively. (b) and (d) are the magnified view of the respective images.

**Figure 3:** Schematic of the mechanism for surface functionalization of filler and bridging between BCT-BZT PVDF polymer through dopamine.

**Figure 4:** Variation of (a-b) dielectric permittivity, (c-d) loss tangent and (e-f) AC conductivity with the frequency of BCT-BZT/PVDF composite films WODA (solid symbol) and WDA (open symbol) modified fillers respectively, (g-i) A comparison of DC dielectric permittivity, dielectric loss tangent, and conductivity with a filler content of BCT-BZT/PVDF composite for WODA and WDA modified fillers.

**Figure 5:** (a-b) Plot of Weibull breakdown strength, (c) variation of breakdown strength and (d) energy storage density of dopamine (WDA), and without dopamine modified (WODA) with filler content in BCT-BZT/PVDF composite.

**Figure 6:** (a) Open-circuit voltage ($V_{oc}$) (b) short-circuit current density ($J_{sc}$) under beating with fist. Inset of (a) shows an enlarged view of a single impact (pressing and release) of fist (c) variation of rectified output voltage ($V_{out}$) and power density with the load resistance. Inset shows the actual flexible device with Al electrode (d) transient response (charging-discharging)

of an external capacitor (10 µF) by the rectified output of 10 wt% WDA composition. Schematic electrical circuits used for the measurement are shown in respective results.

**Figure 1**

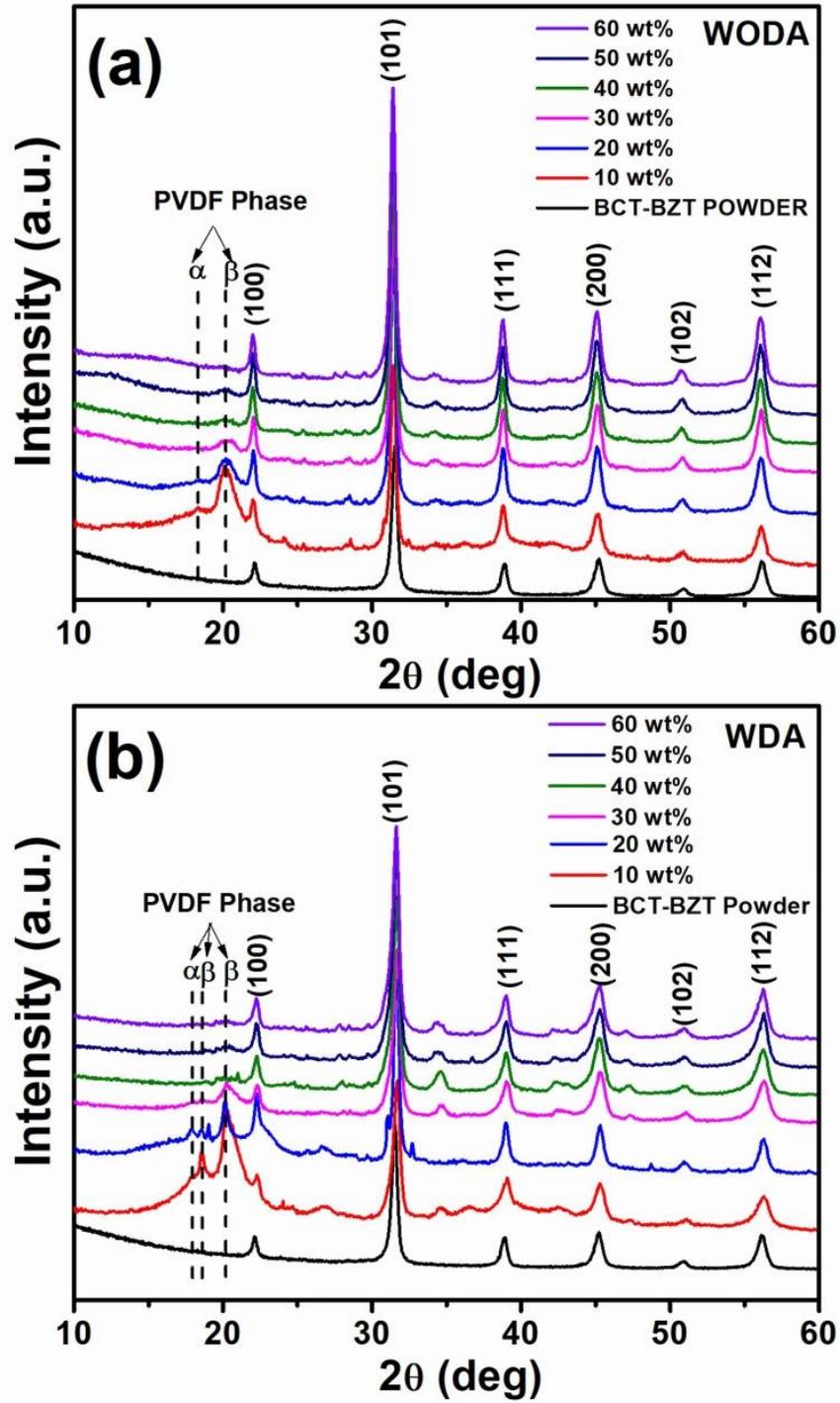

**Figure 2**

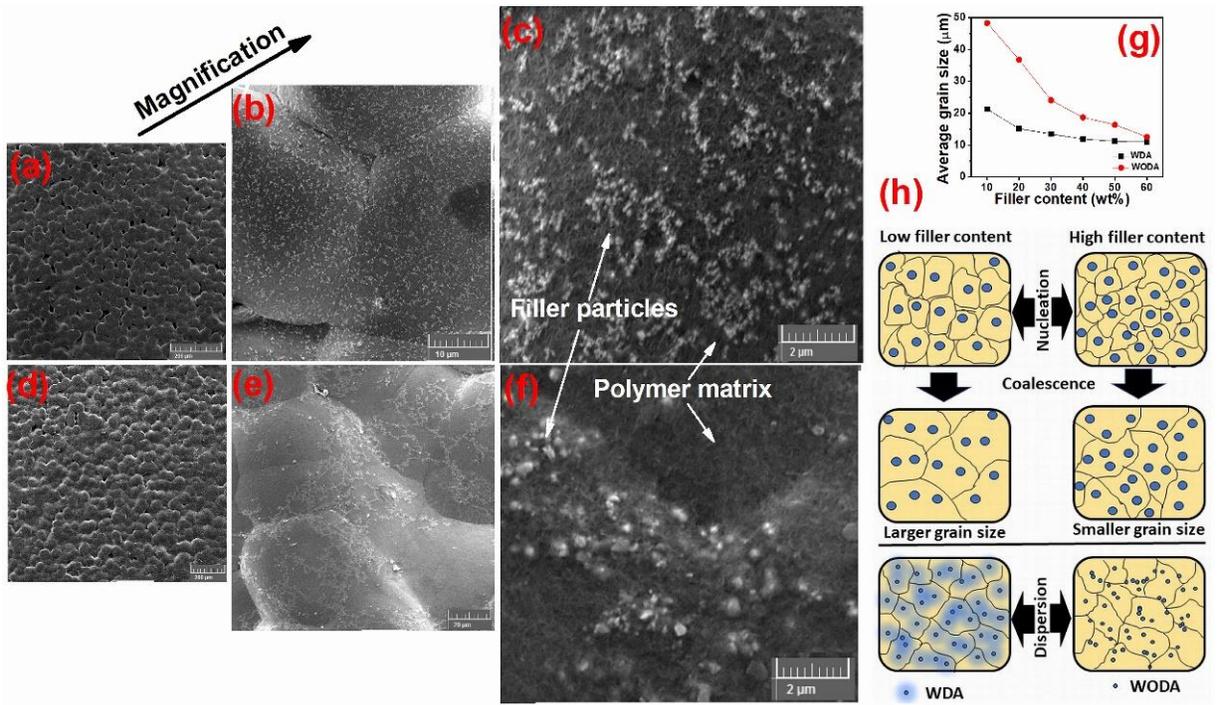

**Figure 3**

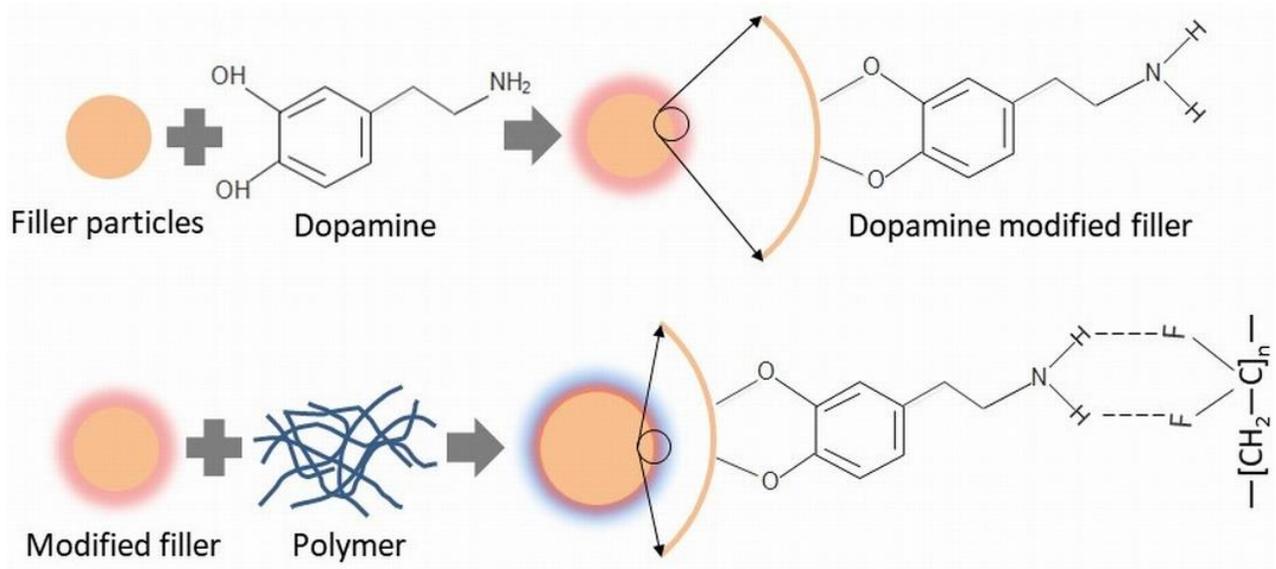

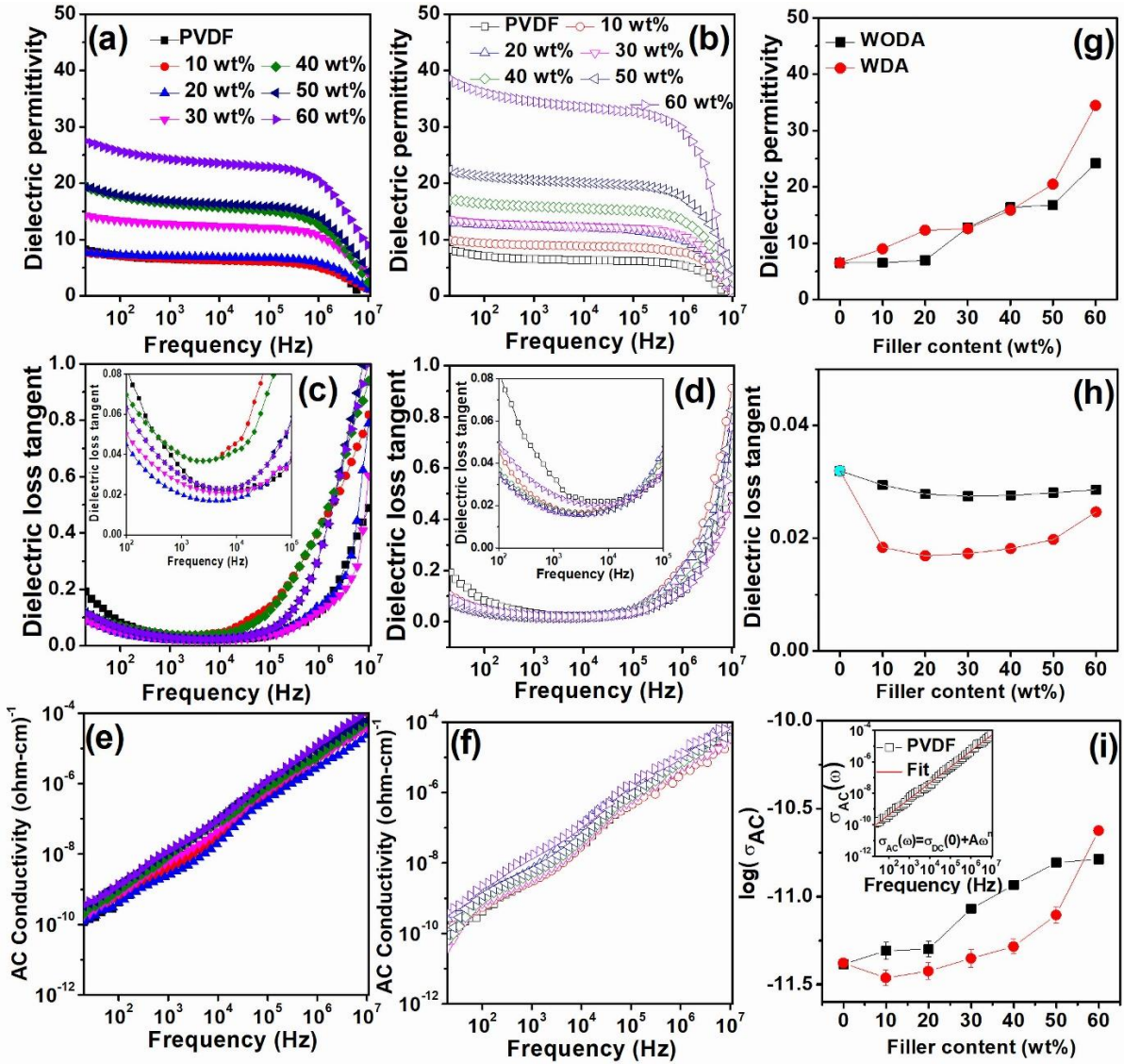

**Figure 5**

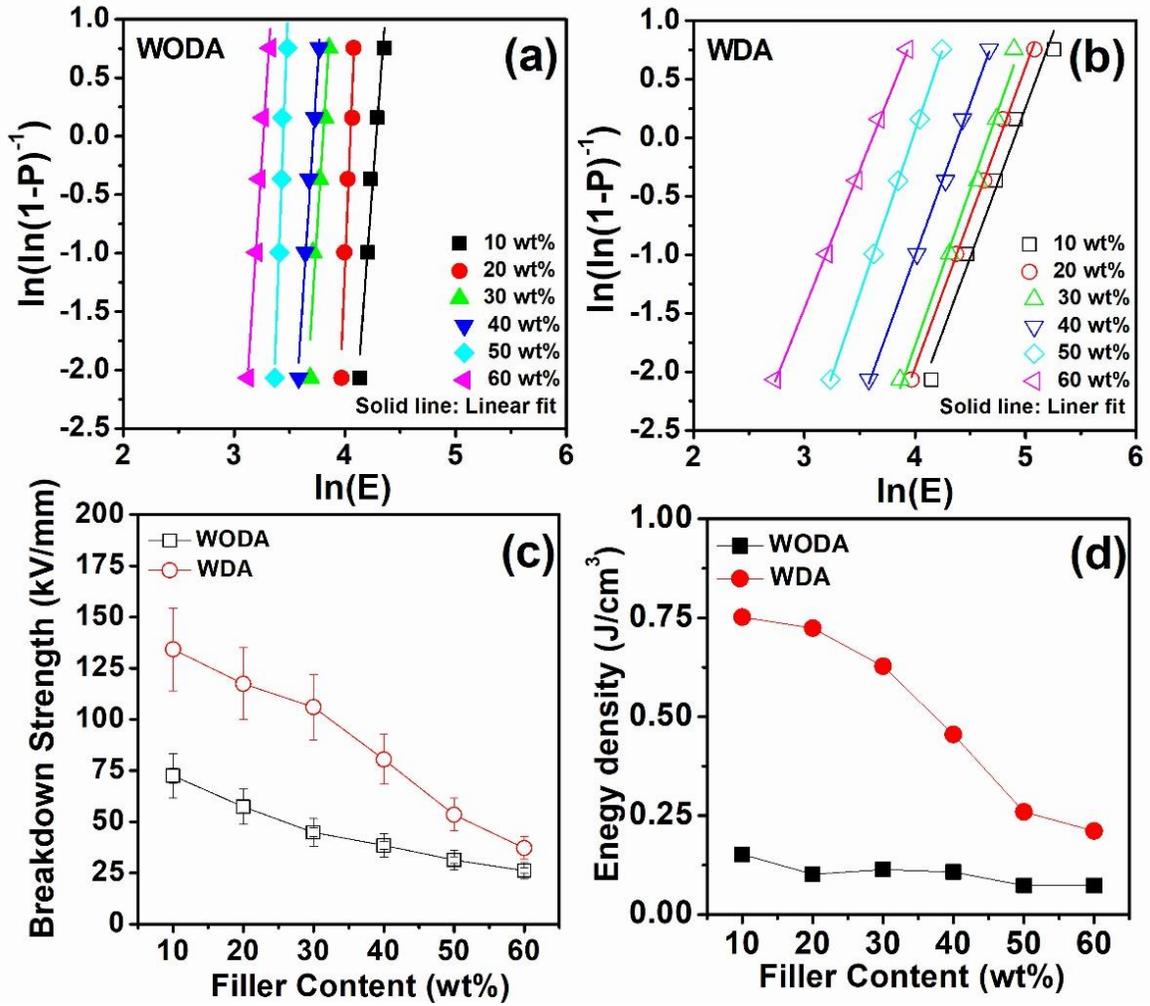

**Figure 6**

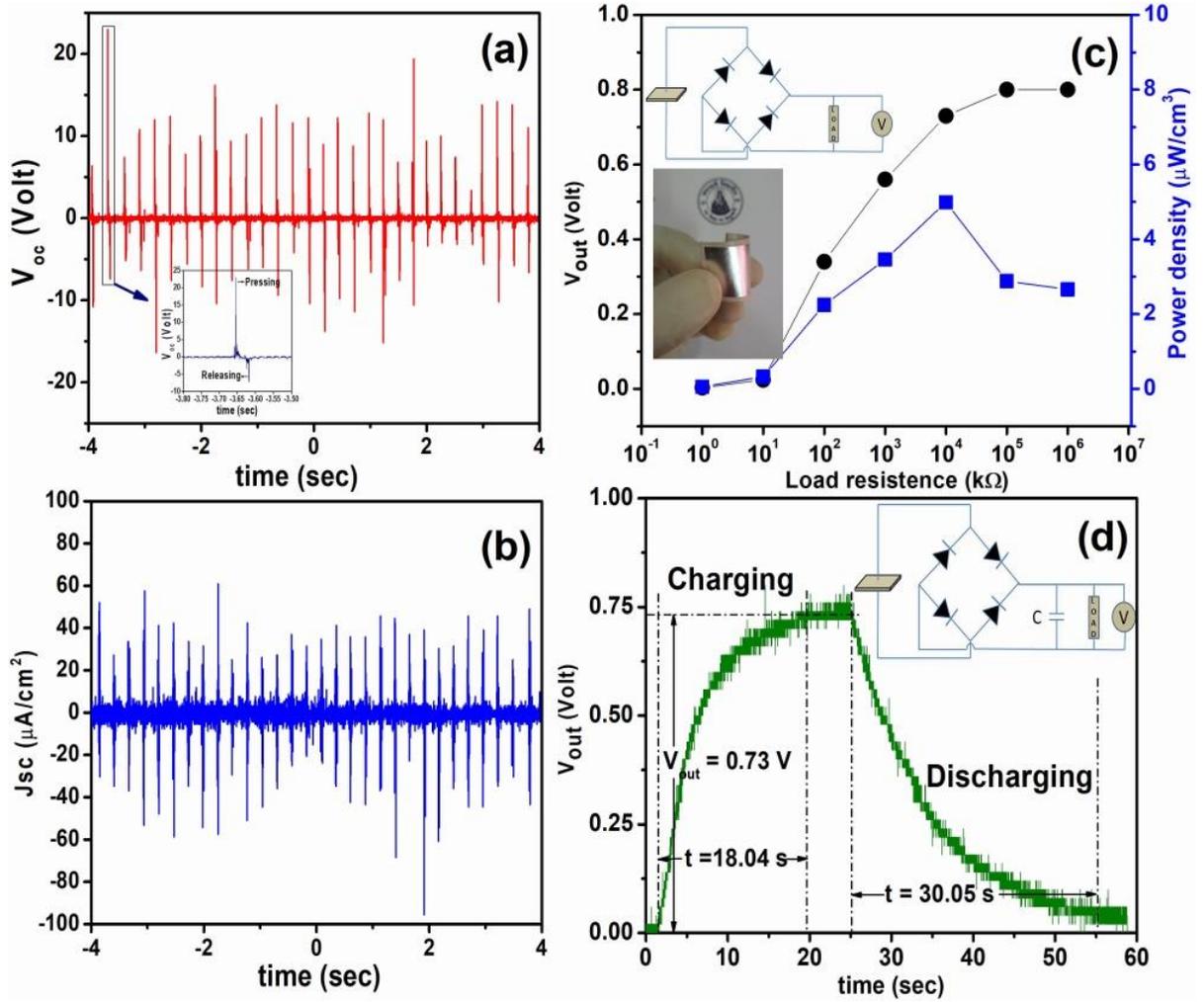